# Computer Application Research based on Chinese Human Resources and Network Information Security Technology Management and Analysis In Chinese Universities

Jun Cui[1, a, *]

1 Solbridge International School of Business, Ph.D., Daejeon, 34613, Republic of Korea.

a jcui228@student.solbridge.ac.kr

* Correspondence: Jun Cui, jcui228@student.solbridge.ac.kr

*Abstract*—This study investigates the current state of computer network security and human resource management within Chinese universities, emphasizing the growing importance of safeguarding digital infrastructures. To support the analysis, interviews were conducted with managers from two leading Chinese cybersecurity firms, and the qualitative data obtained was carefully analyzed to extract key insights and conclusions. The research highlights the critical role of computer technologies in managing cybersecurity and human resources, particularly as universities increasingly rely on digital networks. The study evaluates significant technological innovations in cybersecurity, such as encryption, firewall deployment, and intrusion detection systems, while also examining the human resource management practices that complement these technologies, including risk assessments, policy formulation, and system oversight. Additionally, it explores how big data and artificial intelligence technologies can be integrated with human resource strategies to enhance network security. The findings aim to provide practical solutions to strengthen China's cybersecurity framework, address existing challenges, and enhance overall institutional resilience.

## I. INTRODUCTION

In the modern era of unprecedented global connectivity, the security of network information has emerged as a critical concern, particularly in countries with extensive digital infrastructures like Chinese Universities. Fundamental to maintaining secure and reliable communication are the principles of information integrity, confidentiality, and availability during data transmission across networks. Upholding information integrity is essential for preserving the value of transmitted data and maximizing the efficiency of network information systems. Confidentiality safeguards sensitive user data from unauthorized access, while availability ensures that the transmitted information remains accessible and accurate to its intended recipients [1,2]. Despite significant advancements in cybersecurity measures—including sophisticated encryption techniques and robust firewall protections—network information continues to be vulnerable to tampering, theft, and damage. These risks are further exacerbated by the escalating prevalence of malicious software, such as viruses and hacking activities.

The identified research gap lies in the insufficient convergence of technical solutions and managerial practices to effectively counter these threats in Chinese universities. While a multitude of studies have focused on enhancing cybersecurity through technical means like advanced encryption methods and intrusion detection systems, there is a notable deficiency in research addressing the critical role of human resources management strategies—such as comprehensive risk assessments and the development of robust security policies—in supporting and reinforcing these technologies. Additionally, the rapid evolution and increasing complexity of malicious software, with over 14,000 known variants, have outpaced the creation of comprehensive and adaptable cybersecurity strategies [2,3,4]. This situation highlights the urgent need for a holistic approach that integrates both technological advancements and managerial insights to strengthen cybersecurity frameworks and enhance resilience against both current and emerging threats in Chinese universities.

To address this gap, the present study seeks to answer the following research question:

**How can the integration of advanced cybersecurity technologies and effective human resource management practices enhance the security, integrity, and availability of network information in the face of evolving cyber threats?**

Through exploring this question, the research aims to develop and propose effective human resource strategies that combine technological innovation with managerial frameworks. The goal is to fortify China's cybersecurity infrastructure, effectively respond to ongoing challenges, human resource management and improve overall resilience and stability in the realm of network information security in Chinese universities.

Information integrity ensures that data remains intact during network transmission, playing a critical role in preserving the value of transmitted information and optimizing network functionality [3]. Without it, the utility of human resource information exchange is severely diminished. Confidentiality, a fundamental element of network security, aims to safeguard transmitted data from unauthorized access and misuse, ensuring that staff information remains private and inaccessible to malicious actors. Availability, another crucial aspect of information security, refers to the recipient's ability to access and interpret transmitted data accurately. It also ensures that user data is neither lost nor tampered with during transmission. If user credentials are compromised, availability mechanisms can help recover the lost data [4-7].

Several key threats challenge the security of network information today. First, **network information tampering** occurs when attackers intercept and modify transmitted data. Exploiting the transparency and widespread use of internet technologies, these attackers can alter the data before sending it to the intended recipient, who cannot verify its authenticity, leading to a significant loss of value in the network information (Li & Zhang, 2020). Second, **network information theft and damage** have escalated with the expansion of internet use. As vast quantities of unencrypted information are transmitted daily, cybercriminals target vulnerable data, using sophisticated tools to analyze and crack its structure for illicit gain. Beyond theft, human resource information loss and damage are common, as computer systems are frequently compromised by cyberattacks, resulting in the destruction or corruption of valuable data (Zhou et al., 2019).

The rise of hacking has severely disrupted network security, with hackers often infiltrating systems for financial gain, either by extracting valuable information or sabotaging systems. Such breaches can result in significant losses for victims. Furthermore, malicious software and viruses can infiltrate networks at various stages of human resource information transmission, aiming to corrupt or steal data. Viruses, typically small in size (often under 100 bytes), come in multiple forms such as operating system, shell, invasive, and source code types. They exploit vulnerabilities in devices like floppy disks and terminals to infiltrate computer systems, compromising network stability and security. Some viruses possess self-replicating capabilities, exacerbating their destructive impact by causing system paralysis and information loss [3,4,5,6].

Virus programs are characterized by their stealth and infectiousness, making them difficult to detect and eradicate. When a computer environment of human resource is conducive to virus proliferation, these programs spread rapidly, jeopardizing the security and integrity of, human resources, network information. Currently, over 14,000 virus programs are known, with this number expected to increase as they continue to evolve. The growing threat posed by these programs underscores the urgent need for scientific and technological solutions to effectively curb their spread and mitigate their impact on network information security and human resource management [6,7,8,9].

## II. LITERATURE REVIEW

The growing dependence on digital infrastructures has elevated the importance of network information security in both academic inquiry and practical applications in Chinese universities. With the escalation of cyber threats, a comprehensive understanding of the factors contributing to the protection of human resource information security, integrity, human resources and accessibility is critical. This review explores the theoretical frameworks, key variables, and hypotheses that assess the interplay between technical and managerial practices in enhancing human resource management and network information security in Chinese Universities.

*Key Variables in Network Information Security*

**Information Integrity, Confidentiality, and Availability**. Three essential variables underpin network information security: **information integrity**, **confidentiality**, and **availability**. Information integrity ensures that data remains unaltered or untampered during transmission, safeguarding its accuracy and reliability (Xu et al., 2020). Confidentiality involves the protection of sensitive information from unauthorized access or exposure, serving as a fundamental component of security protocols (Wang et al., 2021). Availability ensures that authorized users can access necessary information, even in cases of security incidents or system disruptions (Li & Zhang, 2020). Despite advancements in cybersecurity technology, these critical factors remain susceptible to cyberattacks. Common risks include data tampering, theft, and system compromise, which jeopardize the overall security framework (Singh & Singh, 2015). Studies have shown that the increasing complexity of cyber threats, such as new viruses and hacking methodologies, exacerbates these vulnerabilities (Zhou et al., 2019).

*Technological Solutions for Network Information Security*
**Encryption, Firewalls, and Intrusion Detection Systems (IDS)**

The literature extensively addresses technical measures that can mitigate network information security risks. **Encryption** plays a pivotal role in ensuring information confidentiality by converting plaintext into an unreadable format, inaccessible to unauthorized individuals (Rizwan et al., 2022). Encryption is especially effective in preventing data breaches during the transmission of sensitive data like user credentials.

**Firewalls**, which act as a protective barrier between trusted internal networks and untrusted external networks, manage the flow of traffic according to pre-established security protocols (Liu et al., 2018). Research consistently shows that firewalls are instrumental in preventing unauthorized access, making them indispensable in modern cybersecurity systems.

**Intrusion Detection Systems (IDS)**, which use both signature-based and anomaly-based detection methods, identify abnormal or malicious activities within a network (Soni et al., 2021). IDS contribute significantly to maintaining data availability by detecting and neutralizing potential threats before they can disrupt network operations.

*Managerial Practices in Cybersecurity*
*Risk Assessment, Policy Formation, and System Monitoring*

While technical solutions provide the foundation for cybersecurity, managerial practices such as **risk assessment**, **policy development**, and **system monitoring** are equally crucial. **Risk assessment** involves identifying, analyzing, and evaluating potential threats to an organization's network infrastructure. Research has found that organizations with robust risk assessment protocols are better prepared to anticipate and mitigate cybersecurity risks (Peltier, 2016).

The development of security **policies** establishes clear guidelines for secure network operations, ensuring that all employees and system users adhere to best practices for data management and incident response (Whitman & Mattord,

2020). Even the most advanced technical measures can be rendered ineffective if not supported by sound policy frameworks, as human error and mismanagement often exacerbate security breaches.

**System monitoring** provides continuous surveillance of network activities, allowing organizations to detect and respond to irregularities and threats in real-time. Effective monitoring minimizes the risk of long-term data loss or system downtime due to cyberattacks (Ghosh et al., 2019).

*The Integration of Technical and Managerial Solutions*

Despite the advances in both technical and managerial cybersecurity measures, a research gap remains in the literature concerning their integration. Many studies focus either on technological innovations or management strategies independently, overlooking the benefits of combining both approaches to improve cybersecurity outcomes. As cyber threats evolve, a holistic approach that integrates both technological and managerial solutions becomes increasingly necessary to address the growing complexities in network security (Xu et al., 2020).

For instance, encryption can secure data from breaches, but it may prove inadequate if not coupled with appropriate management measures. Comprehensive risk assessment and real-time system monitoring are essential to ensure that encryption methods are being properly implemented and regularly updated to respond to emerging threats (Peltier, 2016). Likewise, firewalls and IDS, while effective in mitigating external threats, must be reinforced by security policies that regulate internal network behavior (Liu et al., 2018).

*Theoretical Framework and Hypotheses Development*

The **Diffusion of Innovations (DOI) Theory** (Rogers, 2003) provides a solid theoretical foundation for understanding how cybersecurity technologies are adopted and utilized within organizations. According to DOI, the successful adoption of innovation depends on factors such as perceived usefulness, complexity, and compatibility with existing systems in Chinese universities. In the context of network security, this theory suggests that new technologies—such as encryption and IDS—will yield better results when supported by robust managerial practices that promote their diffusion and integration human resouces within the organization.

Drawing from the DOI theory and the literature reviewed, the following hypotheses are proposed:

- **H1**: Information integrity positively influences the efficacy of cybersecurity technologies.
- **H2**: Information confidentiality enhances network security when supported by strong managerial practices.
- **H3**: Information availability boosts the performance of cybersecurity systems when integrated with risk assessment and system monitoring practices.
- **H4**: The integration of technological and managerial solutions positively impacts overall network security.
- **H5**: The complexity of evolving cyber threats necessitates continuous advancements in both technical and managerial cybersecurity strategies.

This literature review underscores the critical role of both technological solutions and managerial strategies in securing network information. While significant technological advancements—such as encryption, firewalls, and IDS—form the backbone of cybersecurity, they are most effective when complemented by managerial practices, including risk assessments, policy development, human resource management and real-time system monitoring. As cyber threats grow increasingly sophisticated, a coordinated effort combining both technical and managerial elements is essential to maintain the integrity, confidentiality, and availability of network information [3-9].

Future research should further explore the integration of these solutions, thereby advancing the field's understanding of how a comprehensive approach can better protect against the rising tide of cyber threats. Moreover, by addressing this research gap, this study aims to contribute to the ongoing development of a robust cybersecurity framework capable of adapting to the dynamic challenges of the human resource digital age.

The organization of the paper is structured as follows. The introduction provides a comprehensive overview of the study's background, significance, and objectives, while identifying the existing research gap. Following this, the literature review integrates current research on Computer Network information, network security technology, Computer application. The methodology section outlines the research design, data collection methods, and analytical approaches used. This is followed by the results section, which details the outcomes of results. In the discussion, these results are interpreted in relation to the research questions and theoretical framework, with an emphasis on their practical implications. Finally, the paper concludes by summarizing the key findings, addressing the study's limitations, and proposing directions for future research.

III. METHODS AND MATERIALS

In this research, this study employed qualitative interview techniques to gather insights from network security managers across several Chinese Universities. The interviews were designed to explore the current practices, challenges, and strategies in network security management. Specifically, the content of the interviews focused on the following areas:

1. **Current Security Measures**: Managers discussed the technologies and protocols currently implemented within their organizations, including the use of firewalls, encryption, and intrusion detection systems.

2. **Challenges and Vulnerabilities**: Participants shared their experiences regarding common security issues and vulnerabilities they face, such as emerging cyber threats and the effectiveness of their existing security measures.

3. **Management Practices**: The interviews examined the managerial approaches to network security, including risk assessment processes, policy development, and incident response strategies.

4. **Technology Integration**: Insights were gathered on how managers integrate various security

technologies and adapt to new advancements in the field.

5. **Future Trends and Needs**: Managers provided perspectives on future trends in network security and identified areas where further research or technological innovation could be beneficial.

This qualitative approach aimed to provide a deeper understanding of the practical challenges and solutions in network security management and human resource management within the context of Chinese universities[9,10].

In the context of this study, in-depth interviews were conducted with network security managers and human managers from two prominent Chinese technology companies. These interviews provided valuable insights into the practical challenges and strategies associated with network security management (Cui et al., 2024).

**Interview with Human Manager A from TechCorp**

Manager A highlighted the company's robust approach to network security, emphasizing the use of advanced encryption protocols and multi-layered firewalls to safeguard against unauthorized access. According to Manager A, TechCorp employs a proactive risk assessment strategy, which involves regular vulnerability scans and penetration testing to identify potential weaknesses in the network infrastructure. The human resource manager also stressed the importance of integrating security technologies with comprehensive management practices. For instance, TechCorp has developed detailed incident response policies to quickly address and mitigate security breaches. One of the key challenges mentioned was the constant evolution of cyber threats, which necessitates continuous updates to both technical defenses and management strategies. Manager A noted that while current technologies are effective, there is a growing need for adaptive security measures that can respond dynamically to new and emerging threats [12,13,14].

**Interview with Human Manager B from SecureNet**

Manager B provided a different perspective on network security management at SecureNet. This human resource manager focused on the implementation of intrusion detection systems (IDS) and security information and event management (SIEM) tools as central components of the company's security architecture. SecureNet utilizes a combination of signature-based and anomaly-based detection techniques to identify suspicious activities and potential breaches. Manager B emphasized that effective security management extends beyond technology and includes rigorous training programs for employees to recognize and respond to security threats. Additionally, SecureNet places a strong emphasis on policy development, with well-defined protocols for data protection and access control. However, Manager B acknowledged the challenge of balancing security measures with operational efficiency, noting that overly stringent controls can sometimes hinder productivity. The manager expressed the need for continuous innovation in security technologies and management practices to keep pace with the rapidly changing threat landscape.

Both firm interviews reveal that while advanced technologies such as encryption, IDS, and SIEM systems play a crucial role in network security, their effectiveness is significantly enhanced by complementary management practices in Chinese universities.

## IV. RESULTS AND DISCUSSION

The results indicate that the application of identity authentication technologies significantly enhances network security and human resource management by verifying user identities and access rights, reducing the risk of unauthorized access, data tampering, and potential system breaches. The study also highlights that the one-to-one verification system effectively prevents malware intrusions and hacker attacks during data transmission. Additionally, various authentication methods, including secret key verification, biometrics, and trusted objects, offer tailored security solutions based on specific user needs [16,17,18]. Furthermore, the integration of intrusion detection technologies was found to improve real-time threat detection by analyzing abnormal system behaviors, with misuse detection models delivering low false-positive rates. The deployment of firewalls added another layer of security, effectively isolating internal systems from external threats in Chinese universities. The findings also show that antivirus technologies, through targeted virus detection and real-time alerts, play a critical role in maintaining network stability and preventing data loss.

To enhance the control of network information security and human resource management, it is crucial to focus on the development and application of identity authentication technologies. During the transmission of network information, identity authentication plays a critical role in accurately identifying and verifying user identities in Chinese universities. By confirming users' access rights, this technology prevents unauthorized users from accessing the system, mitigating risks of data loss and tampering. This, in turn, creates a more secure and stable environment for computer systems. Furthermore, identity authentication operates by comparing and verifying parameters during usage, establishing a trust verification mechanism between users and systems, thus enhancing overall network security. The "one-to-one" verification structure of identity authentication is particularly effective in preventing virus intrusions and hacker attacks during information transmission [11].

As research into authentication technologies progresses, various methods have emerged, such as secret key verification, trusted objects, and biometrics. Each method serves distinct functions, allowing users to select the most suitable technology based on their specific needs and the characteristics of the network. This flexibility helps prevent issues like data loss or tampering. In addition, the integration of intrusion detection technology further strengthens security by detecting unauthorized access or abnormal behavior in real-time. Intrusion detection typically follows three steps: information collection, human information analysis, and response processing. Through the collection of security logs and system data, potential intrusions are analyzed, allowing for immediate response measures [7-10].

Currently, the two primary intrusion detection models are misuse detection and anomaly detection. Misuse detection is effective at analyzing attack types and generating detailed reports with low false-positive rates. Conversely, anomaly detection focuses on identifying deviations from normal operations, though it may exhibit higher false-positive rates in some scenarios. Another critical component of network security is the firewall. Likewise, Firewalls act as barriers against unauthorized access, isolating external threats and safeguarding internal system operations [6].

Viruses also pose a significant threat to network information security and human resource management. The application of antivirus technologies is essential for enhancing the overall security framework, as it enables the detection and elimination of virus programs [8]. Human resources users can select from various virus detection tools tailored to their specific security needs, either targeting specific virus programs or conducting comprehensive system scans in Chinese universities. Additionally, the configuration of a virus protection system within the network allows for real-time alerts in the event of virus infiltration, providing timely protection against potential data loss [9,10,11]. To address these challenges, it is vital for organizations to invest in the human resource development of advanced security technologies. This will not only improve current network security management practices but also foster long-term advancements in the field of computer science. In this regard, government support and adequate funding are essential for facilitating research and innovation in information security technologies, including encryption, authentication, and network monitoring [15-18].

Additionally, addressing the shortage of skilled professionals in the field of network security is imperative. Collaboration with universities to offer specialized programs in information security can expand the pool of qualified professionals in Chinese universities. Enhancing salaries and benefits for IT professionals, coupled with personalized training programs, can help attract and retain high-caliber talent. Introducing reward systems and fostering a sense of responsibility among network security personnel can also contribute to improved performance in this domain [19].

Lastly, improving the existing security management frameworks is critical. Relevant authorities should develop integrated security control models and implement comprehensive management systems that align with current network security needs in Chinese universities [22]. Government agencies must also play a proactive role by formulating policies and regulations that address emerging security risks. Encouraging enterprises to adopt standardized security practices and regularly perform virus detection and prevention measures will further strengthen the overall network security environment, reducing vulnerabilities and ensuring long-term protection against cyber threats [23].

## V. CONCLUSIONS

This study highlights the crucial role of advanced identity authentication and intrusion detection technologies in fortifying human resources and network information security. The findings demonstrate that effective user verification mechanisms, including biometrics and secret key systems, significantly mitigate risks associated with unauthorized access and data tampering. Intrusion detection technologies, particularly misuse detection models, enhance real-time threat identification with minimal false positives. The integration of firewalls and antivirus solutions further bolsters network defenses by isolating threats and preventing malware infections. To address evolving cybersecurity challenges, it is essential for organizations to adopt a multifaceted security approach that combines technological innovations with robust management practices in Chinese universities. Continuous investment in cutting-edge technologies and the development of skilled personnel are vital for maintaining a secure network environment. Future research should explore the synergy between these technologies and organizational strategies to build resilient security frameworks that can effectively counteract emerging cyber threats (Sun et al., 2021).

These findings suggest that identity authentication and intrusion detection technologies are essential for strengthening network information security in a rapidly evolving human resource and digital environment (Sun et al., 2021). While this study provides valuable insights into the integration of identity authentication and intrusion detection technologies for enhancing human resource management and network information security, it is important to acknowledge several limitations that may affect the generalizability and applicability of the findings. Firstly, the research primarily focuses on technological solutions and their management aspects within the context of network security. This approach, while comprehensive, may overlook other critical factors such as organizational culture, user behavior, and the broader socio-economic context that could influence the effectiveness of these security measures. For example, the adoption and implementation of advanced security technologies may be hindered by organizational resistance or lack of awareness among staff, which could undermine the overall security posture [13-15].

Secondly, the study's scope is limited to specific technologies and management practices currently prevalent in the field. As cybersecurity is a rapidly evolving domain, new threats and technological advancements continually emerge. The static nature of the technologies examined may not fully account for future developments, potentially limiting the relevance of the findings over time. Additionally, the study does not explore the effectiveness of integrating emerging technologies and human resources, such as artificial intelligence and machine learning, which are increasingly being utilized in cybersecurity to enhance threat detection and response [21-24].

Another limitation is the focus on general application scenarios without considering industry-specific contexts. Different sectors, such as finance, healthcare, and government, face unique cybersecurity challenges and regulatory requirements. The strategies and technologies that work well in one sector may not be directly applicable to another. Therefore, the findings may not be uniformly applicable across various industries, which could affect the practical implementation of the proposed security measures.

Furthermore, the research is based on a review of existing literature and theoretical frameworks, which may not always capture the practical challenges and nuances faced by organizations in real-world scenarios. The effectiveness of the recommended technologies and practices may vary depending on factors such as organizational size, complexity, and existing infrastructure. This gap highlights the need for empirical studies and case analyses to validate and refine the proposed approaches in diverse organizational settings. Future work should consider the development of comprehensive security frameworks that integrate technological solutions with organizational strategies, policies, and processes. This holistic approach can address the dynamic nature of cybersecurity threats and ensure that organizations are better equipped to handle evolving risks and challenges in Chinese universities.

In conclusion, while this study offers important contributions to the field of network information security, addressing its limitations and pursuing future research directions can lead to more robust and adaptive security solutions in Chinese universities. By expanding the scope of research to include organizational, industry-specific, and technological advancements, the field can continue to advance and improve the resilience of network information security, human resource management. Consequently, this study also examines the current landscape of network security and human resource management in Chinese universities, emphasizing the critical role that computer technologies play in cybersecurity efforts. With the increasing dependence on digital infrastructures, safeguarding sensitive information has become a top priority. To support the analysis, interviews with managers from two prominent Chinese cybersecurity firms were conducted, providing valuable insights. The qualitative analysis of these interviews highlights significant findings regarding both technological innovations and management strategies. Key technologies such as encryption, firewalls, and intrusion detection systems are evaluated in the context of university cybersecurity. Additionally, the study explores the human resource management practices that support these technologies, including risk assessment, policy formulation, and system oversight. By integrating advanced technologies with human resource strategies, particularly through enterprise technology and artificial intelligence, this research proposes actionable solutions to enhance the cybersecurity infrastructure in Chinese universities. It aims to address existing challenges, bolster resilience, and strengthen the overall security framework in the academic environment.

ACKNOWLEDGMENT

This research has been supported/partially supported Solbridge International School of Business, Thanks to all contributors.

ORCID

Jun Cui  https://orcid.org/0009-0002-9693-9145